\documentclass[graybox]{svmult}
\usepackage[utf8]{inputenc}

\usepackage{type1cm}        %
\usepackage{makeidx}         %
\usepackage{graphicx}        %
\usepackage{multicol}        %
\usepackage[bottom]{footmisc}%

\usepackage[colorlinks = true,
            linkcolor = blue,
            urlcolor  = blue,
            citecolor = blue,
            anchorcolor = blue]{hyperref}

\usepackage{newtxtext}       %
\usepackage{newtxmath}       %

\makeindex             %

\usepackage{fancyhdr}
\usepackage{ragged2e}
\fancypagestyle{bchap}{%
\fancyhead[L,R] {}

\fancyfoot[L] {\footnotesize \justifying \noindent Steven Van Vaerenbergh and Adrián Pérez-Suay. A Classification of Artificial Intelligence Systems for Mathematics Education. In P. R. Richard, P. Vélez, S. Van Vaerenbergh (Eds.). {\em Mathematics Education in the Age of Artificial Intelligence.} Springer Nature, in press.}
\fancyfoot[C] {}}

\begin{document}

\title*{A Classification of Artificial Intelligence Systems for Mathematics Education}
\author{Steven Van Vaerenbergh and Adrián Pérez-Suay}
\institute{Steven Van Vaerenbergh \at Departamento de Matemáticas, Estadística y Computación, Universidad de Cantabria, Avda. de los Castros 48, 39005 Santander, Spain, \email{steven.vanvaerenbergh@unican.es}
\and Adrián Pérez-Suay \at Departament de Didàctica de la Matemàtica, Universitat de València, Avda. Tarongers, 4, 46022 València, Spain, \email{adrian.perez@uv.es}}
\maketitle

\abstract{This chapter provides an overview of the different Artificial Intelligence (AI) systems that are being used in contemporary digital tools for Mathematics Education (ME). It is aimed at researchers in AI and Machine Learning (ML), for whom we shed some light on the specific technologies that are being used in educational applications; and at researchers in ME, for whom we clarify: i) what the possibilities of the current AI technologies are, ii) what is still out of reach and iii) what is to be expected in the near future. We start our analysis by establishing a high-level taxonomy of AI tools that are found as components in digital ME applications. Then, we describe in detail how these AI tools, and in particular ML, are being used in two key applications, specifically AI-based calculators and intelligent tutoring systems. We finish the chapter with a discussion about student modeling systems and their relationship to artificial general intelligence.
}

\section{Introduction}

Artificial intelligence (AI) has a long history, starting from observations by the early philosophers that a reasoning mind works in some ways like a machine. For AI to become a formal science, however, several advances in the mathematical formalization of fields such as logic, computation and probability theory were required \cite{Russell2009}. Interestingly, the relationship between mathematics and AI is not unilateral, as AI, in turn, serves the field of mathematics in several ways. In particular, AI powers many computer-based tools that are used to enhance the learning and teaching of mathematics, several of which are the topic of discussion of this chapter.

The close relationship between AI and Mathematics Education (ME) dates back at least to the 1970s, and it has been discussed thoroughly in the scientific literature \cite{schoenfeld1985artificial,wenger1987artificial,Balacheff1994}. One could list several parallels between both fields, for instance that they are both concerned with constructing sound \emph{reasoning} based on the use of logic. Indeed, in ME, developing mathematical reasoning skills is an important educational goal, while many AI systems are designed to perform reasoning tasks in an automated manner. 
Also, modern AI techniques involve the concepts of \emph{teaching} and \emph{learning}, as some systems are required to learn models and concepts, either in an autonomous manner or supervised through some form of instruction\footnote{We will not enter into details regarding the similarities and differences in learning for AI and ME, as that discussion is slightly outside of the scope of this chapter.}. Nevertheless, while these parallels exist, humans and machines clearly carry out these tasks in completely different ways. After all, as noted by Schoenfeld, ``AI's perspective is severely distorted by the engineering perspective, and extrapolations to human performance can be dangerous'' \cite[p. 184]{schoenfeld1985artificial}.

Before continuing, we will take a closer look at what exactly AI is, including its subfield, machine learning.

\subsection{Artificial intelligence and machine learning}
The literature contains many different definitions of AI, though they are all much related. Generally speaking, AI aims to create machines capable of solving problems that appear hard to the eyes of a human observer. Such problems may be related solely to thought processes and reasoning capabilities, or they may refer to exhibiting a certain behavior that strikes as intelligent \cite{Russell2009}.

Historically, several simple AI programs have been designed using a set of predefined rules, which can often be represented internally as a decision tree model. For instance, a program for natural language understanding may look if certain words are present in a phrase, and combine the results through some set of fixed rules to determine the sentiment of a text. And in early computer vision systems, results were obtained by calculating hand-engineered \emph{features} of pixels and their neighborhoods, after which these features were compared to the surrounding features using predefined rules. However, as soon as one tries to build a system with an advanced comprehension of natural language or photographic imagery, a very complex set of such internal rules is required, exceeding largely what can be manually designed by a human expert. To deal with this complexity, an \emph{automated} design process for the set of internal rules is required, better known as \emph{machine learning}.

Formally, machine learning (ML) is a subfield of AI that follows a paradigm known as \emph{learning from examples}, in which a system is given practical examples of a concept or behavior to be learned, after which it develops an internal representation that allows its own output to be consistent with the set of given examples \cite{wenger1987artificial}. The concept of ML is perhaps best summarized by Tom Mitchell, who wrote it is ``the field that is concerned with the question of how to construct computer programs that automatically improve with experience'' \cite{mitchell1997machine}. This definition highlights the three properties that any ML system should hold: 1) its learning is \emph{automated}, as in a computer program that does not require human intervention; 2) the performance can be measured, which allows the system to measure \emph{improvement}; and 3) learning is based on receiving examples (the \emph{experience}). In summary, as more examples are processed, a well-designed ML system guarantees that its performance will improve.

The concept of computer programs that learn from examples has been around as long as AI itself, notably as early as Alan Turing's vision in the 1950s \cite{Russell2009}. Nevertheless, it was not until the middle of the 1990s that a solid foundation for ML was established by Vladimir Vapnik in his seminal work on statistical learning theory \cite{Vapnik1995}. At the same time, the techniques of neural networks and support vector machines gained popularity as they were being applied in real-world applications, though only for simple tasks, as compared to today's standards. The latest chapter in the history of ML started around 2012, when it became clear that neural networks could solve tasks that were much more demanding than previously achieved, by making them larger and feeding them much more examples \cite{krizhevsky2012imagenet}. These neural networks consist of multiple layers of neurons, leading to the name \emph{deep learning}, where each layer contains thousands or even millions of parameters.
They are responsible for many of the AI applications that people interact with at the time of writing, notably in image and voice recognition, and natural language understanding.

In this chapter, we aim to analyze the uses of AI, and in particular ML, in ME. The chapter is structured as follows. To fix ideas, in Section \ref{sec:examples} we briefly discuss some contemporary AI-based tools that are being used by mathematical learners. Based on these examples, we introduce a high-level taxonomy of AI systems that serve as building blocks for tools in ME, in Section \ref{sec:taxonomy}. We then show, in Section \ref{sec:analysis}, how the proposed taxonomy allows us to provide an in-depth analysis of the different AI systems currently used in some ME tools. We finish with a discussion on future research required to build complete student modeling systems, in Section \ref{sec:studentmodeling}, and conclusions, in Section \ref{sec:conclusions}.

\section{A glimpse of the present}
\label{sec:examples}

The following are two examples of AI-based tools for mathematics education. These examples feature some representative AI techniques that will be at the basis of the taxonomy we propose later, after which we will revisit them for a more in-depth analysis.

\subsection{A new breed of calculators}
\label{sec:reasonators}

In 2014, a mobile phone application called Photomath\footnote{\url{https://photomath.app/}} was released that quickly became very popular among schoolchildren (and less so among teachers) \cite{webel2015teaching}. It allows the user to point a phone camera at any equation in a textbook and instantly obtain the solution, including the detailed steps of reasoning. If the user wishes so, he or she can even request an alternative sequence of steps for the solution. While the first version of the software would only work on pictures of clean, textbook equations, the technology was later upgraded to recognize handwriting as well. Several other apps have followed suit since, notably Google's Socratic\footnote{\url{https://socratic.org/}} and Microsoft Math Solver\footnote{\url{https://math.microsoft.com/}}.

The emergence of these tools, known as ``camera calculators'', can be mainly attributed to advances in image recognition technology, and in particular to optical character recognition (OCR) algorithms based on deep learning. Once the OCR algorithm has translated the picture into a mathematical equation, standard equation solvers can be used to obtain a solution. The third and last stage of the application consists in explaining the solution to the user, for instance, by means of a sequence of steps. In the case of the solution of a linear equation, this stage is resolved algorithmically, requiring little AI. Nevertheless, it is often possible to find a shorter or more intuitive solution by thinking strategically (see, for instance, the example in \cite[p. 4]{webel2015teaching}). For an AI to do so, it would require capabilities of mimicking human intuition or exploring creative strategies. We will discuss these capabilities later, in Section \ref{sec:exploration}.

Interestingly, these apps have reignited the discussion on the appropriate use of tools in mathematics education, reminiscent of the controversy on the use of pocket calculators that started several decades ago \cite{webel2015teaching}. Indeed, they could be seen as examples of a new generation of ``smarter'' calculators, that limit the number of actions and calculations that the user must perform to reach a solution. Such new, smarter calculators are not limited to equation solvers only, as they can be found in other fields as well. For instance, as of version 5, the popular dynamic geometry software (DGS) GeoGebra\footnote{\url{https://www.geogebra.org/}} includes a set of automated reasoning tools that allow the rigorous mathematical verification and automatic discovery of general propositions about Euclidean geometry figures built by the user \cite{kovacs2021towards}. Rather than merely automating calculations, as common digital calculators do, these tools allow to \emph{automate the reasoning}, to a certain extent.

\subsection{Blueprint of a data-driven intelligent tutoring system}
\label{sec:blueprintITS}

Our second example concerns an interactive tool for learning and teaching mathematics. In particular, in the following, we describe a hypothetical interaction between a student and an intelligent tutoring system (ITS).

Hypatia, a student, logs onto the system through her laptop and she starts reading a challenge proposed by the system. This time, the challenge consists in solving an integral equation. Hypatia is not sure how to start, and she spends a few minutes scratching calculations on her notepad. The system, after checking her profile in the database, infers that she needs help, and offers a hint on screen. Hypatia now knows how to proceed and advances a few steps towards the solution. However, some steps later, she makes a mistake in a substitution. The system immediately notices the mistake and identifies it as a common error (a ``bug''). Through the visual interface, the system tells Hypatia to check if there were any mistakes in the last step. She reviews her calculations and quickly corrects the mistake. The system encourages her for spotting the error, and she continues to solve the exercise successfully. At this point, the system shows her a summary of the solution and reminds her of the hints she was given. She can then choose to review any of the steps and their explanation, or continue to the next problem. If she chooses to continue, the system will present her a problem that has been designed specifically to advance along her personalized learning path.

Before Hypatia started using this ITS, the system's database already contained the interactions of many other students. By using data mining techniques, it was able to identify a number of ``stereotype'' student profiles. The first time Hypatia interacted with the ITS, the system's AI analyzed her initial actions to build an initial profile for her, based on one of the stereotype profiles. As she now performs different problem solving sessions, the system adds more of her interactions to its database, which allows it to identify behavioral patterns and build a more refined student model for her. This, in turn, allows the system to personalize her learning path and to offer her more relevant feedback when she encounters difficulties.

Certainly, the above-described example is not purely hypothetical, but based on real ITS that are used in practice today. We will return to this example later on.

\section{A taxonomy of AI techniques for mathematics education}
\label{sec:taxonomy}

We now propose a taxonomy of AI techniques that are used in digital tools for ME. The taxonomy consists of four categories that span the entire range of such AI systems. 
While each of the categories is motivated by some aspect of the previous examples, we include a more comprehensive list of particular cases from the literature for each of them. Furthermore, we shed some light on the current technological capabilities of these AI systems.

\subsection{Information extractors}
\label{sec:extractors}

\begin{figure}[h]
\centering
\includegraphics[scale=0.2]{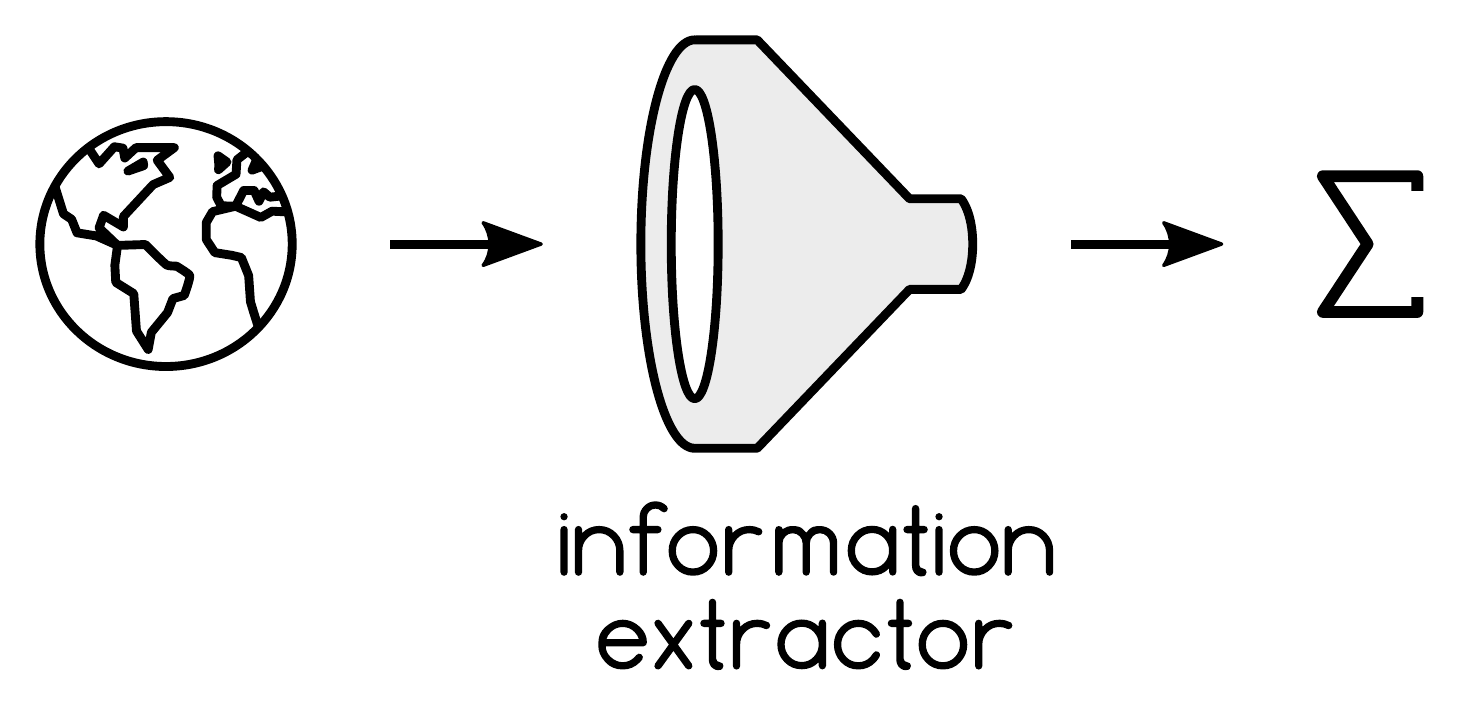}
\caption{Representation of an information extractor. The globe represents observations from the real world, and the summation sign represents mathematical information.}\label{fig:icon_information_extractor}
\end{figure}

We use the term \emph{information extractors} to refer to AI technologies that take observations from the real world and translate them into a mathematical representation (Fig. \ref{fig:icon_information_extractor}). A classic example in this category consists in parsing the text of algebraic word problems into equations \cite{koncel2015parsing}. More advanced information extractors can operate on digitized data from a sensor, such as a camera or a microphone, to which they apply an AI algorithm to extract computer-interpretable mathematical information.

An example of information extractors that operate on visual data was given in Section \ref{sec:reasonators}, where the initial stage of the described camera calculator translates a picture into a mathematical equation. %
The AI required to perform OCR in these information extractors operates in two steps: First, it employs a convolutional neural network (CNN) to recognize individual objects in an image. In essence, a CNN is a particular type of artificial neural network that is capable of processing spatial information present in neighborhoods of pixels by applying (and learning) digital filters \cite{krizhevsky2012imagenet}. Then, the individually recognized objects are transformed into a sequence, which was traditionally performed by techniques such as Hidden Markov Models \cite{rabiner1986introduction}, but is now implemented as neural-network based techniques such as Long Short-Term Memory networks \cite{hochreiter1997long} and transformers \cite{vaswani2017attention}.

Visual information extractors are not only used to digitize algebraic equations, but can also be used to extract other types of mathematical information from the real world. For instance, in the MonuMAI project \cite{lamas2021monumai}, extractors based on CNN are used to obtain geometrical information from pictures of monuments. And some camera calculators, such as Socratic, allow to take pictures of word problems, which are transformed to text, interpreted, and converted into a mathematical representation.

Finally, sensor data from a student may be used to extract information for an ITS (see Section \ref{sec:its}). In particular, these systems may require information about the student's state of mind during the resolution of a mathematical problem. In this category we encounter AI techniques for facial expression recognition \cite{li2020deep}, speech emotion recognition \cite{ fayek2017evaluating}, and mood sensing through electrodermal activity \cite{kajasilta2019measuring}.

\subsection{Reasoning engines}

\begin{figure}[h]
\centering
\includegraphics[scale=0.2]{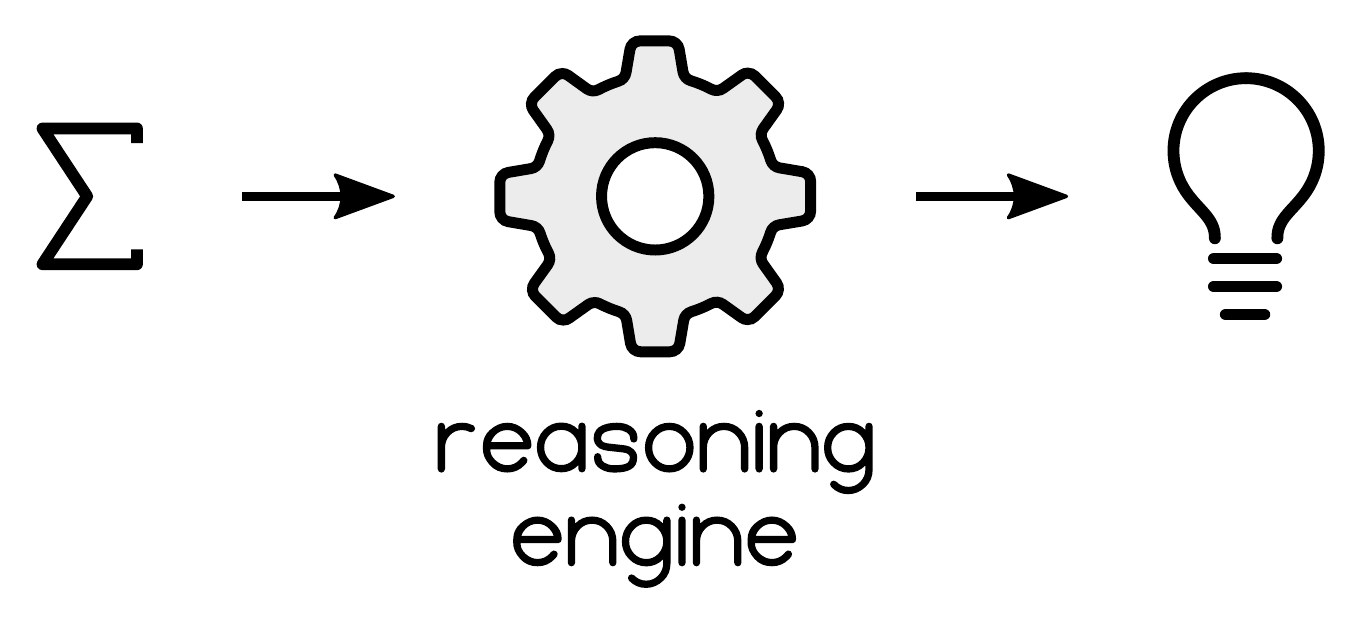}
\caption{Representation of a reasoning engine. It receives a mathematical problem as an input, and outputs the corresponding solution.}\label{fig:icon_reasoning_engine}
\end{figure}

In software engineering, a \emph{reasoning engine} is a computer program that is capable of inferring logical consequences from a set of axioms found in a knowledge base, by following a set of predefined rules \cite{furht2008encyclopedia}. For the current context of mathematics education, we employ a broader definition of reasoning engines that includes all software systems that are capable of automatically solving a mathematically formulated problem (Fig. \ref{fig:icon_reasoning_engine}). A very simple such system consists of an equation solver, whose action is limited to transforming the (set of) equations into their canonical form and applying the formula or the algorithm to solve them \cite{Arnau2013}. Several types of more sophisticated reasoning engines exist in the mathematical research literature, for instance \emph{automated theorem provers} (ATP), whose aim is to verify and generate proofs of mathematical theorems \cite{loveland1978automated}. While proof verification is a simple mechanical process that only requires checking the correctness of each individual step, proof generation is a much harder problem, as it requires searching through a combinatorial explosion of possible steps in the proof sequence.

A novel contribution in the development of reasoning engines lies in the use of ML techniques, which has been fueled by the success of deep learning in pattern matching problems \cite{krizhevsky2012imagenet}. These techniques follow the standard ML paradigm that requires a set of training examples: The ML algorithm, typically a deep neural network, learns a model in order to explain as much of the training examples as possible. The learned model is completely data driven, without any hard rules or logic programmed into it.

ML algorithms could improve current ATP techniques by encoding human provers' intuitions and predicting the best next step in a proof \cite{gauthier2015premise,loos2017deep,schon2019using}. Furthermore, neural networks for natural language processing are being used to train machines to solve word problems and to perform symbolic reasoning, yielding currently some limited but promising results. For instance, Saxton et al. \cite{saxton2019analysing} generated a data set of two million example problems from different areas of mathematics and their respective solutions. Several neural network models were trained on these data and, in general, a moderate performance was obtained, depending on the problem type. Deep learning is also being used to solve differential equations \cite{arabshahi2018towards,lample2019deep}, perform symbolic reasoning \cite{lee2020mathematical}, and solve word problems \cite{wang2017deep,wang2018mathdqn}. Note that these methods typically operate on text data and they perform the action of the information extractor and the reasoning engine using a single AI. Finally, in the ML community there is a growing interest in automating abstract reasoning. Research in this area currently focuses on solving visual IQ tests, such as variants of Raven's Progressive Matrices \cite{barrett2018measuring,chollet2019measure}, and causal inference, which deals with explaining cause-effect relations, for instance from a statistical point of view \cite{Pearl2019}.

\subsection{Explainers}

While reasoning engines can solve mathematical problems and generate correct proofs, they do not necessarily produce results that can be read by a human. Sometimes this is simply not needed, for instance when an ATP is used in research to verify a theorem that requires a long and complex proof, prone to human errors. But in a different context, for instance that of the mathematical learner, it becomes important to have proofs that are understandable \cite{ganesalingam2017fully}.

\begin{figure}[h]
\centering
\includegraphics[scale=0.2]{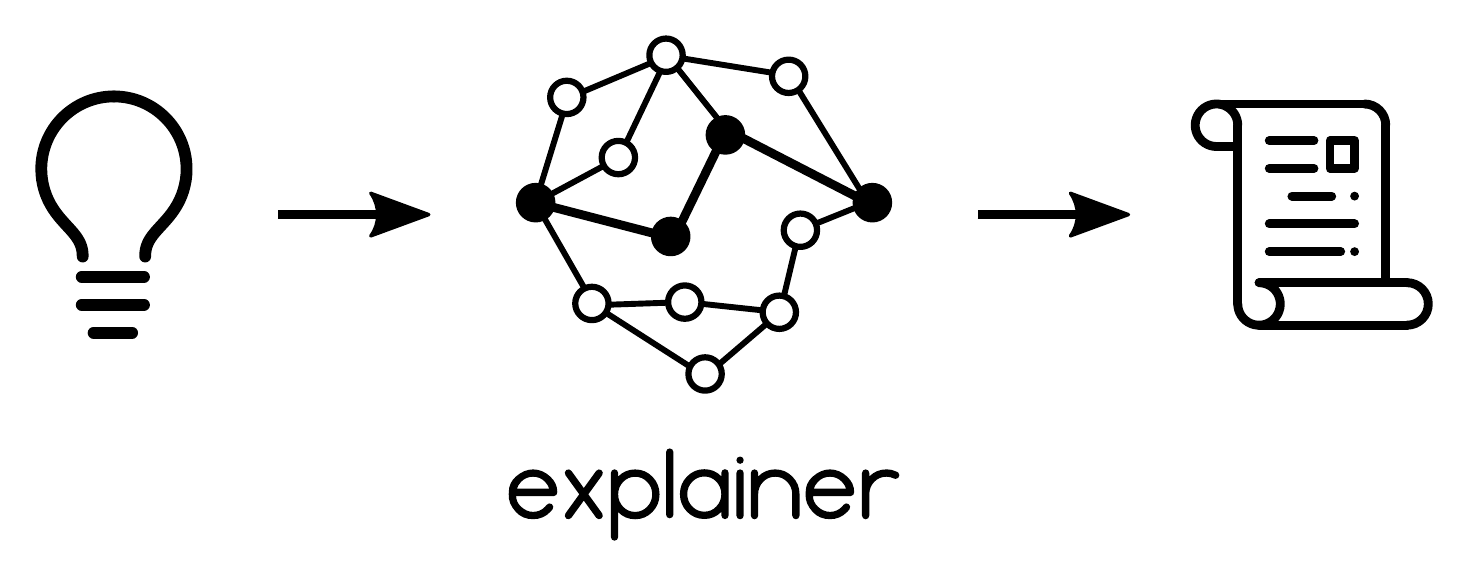}
\caption{Representation of a post-hoc explainer. It translates a machine-code solution into a sequence of logical, human-readable steps.}\label{fig:icon_explainer}
\end{figure}

In the AI community, interest in explainable methods has recently surged. Part of this interest is due to legal reasons, as some administrations demand that decisions taken by an AI model on personal data be accompanied by a human-understandable explanation \cite{How2019}. While some early AI systems generated models that could easily be interpreted, modern AI techniques, especially deep learning systems, involve opaque decision systems. These algorithms operate in enormous parametric spaces with millions of parameters, rendering them effectively black-box methods whose decisions cannot be interpreted. To solve this issue, the research field of \emph{explainable AI} is concerned with developing AI methods that produce interpretable models and interpretable decisions \cite{adadi2018peeking,molnar2019,arrieta2020explainable}. We will refer to AI methods that produce understandable explanations as \emph{explainers} (Fig. \ref{fig:icon_explainer}).

From a technical point of view, there exist two types of explainers. The first type are modules that can be added onto existing, opaque AI systems. They perform what is called \emph{post-hoc explainability}, and may do so for instance by approximating the complex model with a simpler, interpretable one. Some different post-hoc explainability approaches are illustrated in \cite[Fig. 4]{arrieta2020explainable}. The second type of explainable AI consists of models that are interpretable by design. Under our terminology, these correspond to reasoning engines that are restricted to only producing interpretable solutions. Of these two types, the former has the advantage that it does not require replacing the entire reasoning engine, which is usually hard to design and train in the first place.

In the field of ME, explainers have been built principally for solving math equations step by step, for instance in the open source project mathsteps\footnote{\url{https://github.com/google/mathsteps}}. In ATP, on the other hand, explainability is a fairly new research line. In order to apply a post-hoc explainer onto an ATP, it might be necessary to construct an ATP based purely on logic, though, as \cite{fu2019robot} notes, ``while logic methods proposed have always been the dream of mankind, their applications are limited due to the massive search space''. One case in point is found in DGS, where geometric automated theorem provers (GATP) are now being integrated \cite{quaresma2020automated}. State-of-the-art GATP are based on algebraic methods, and their results cannot be translated into human-readable proofs \cite{quaresma2020automated,kovacs2020geogebra}. For this reason, explainability is to be introduced in ATP by designing ATP techniques that are transparent by design \cite{ganesalingam2017fully,How2019}. In the case of GATP, this approach is currently very limited, as discussed in \cite{font2018improving}.

\subsection{Data-driven modeling}
\label{sec:data_driven_modeling}

\begin{figure}[h]
\centering
\includegraphics[scale=0.2]{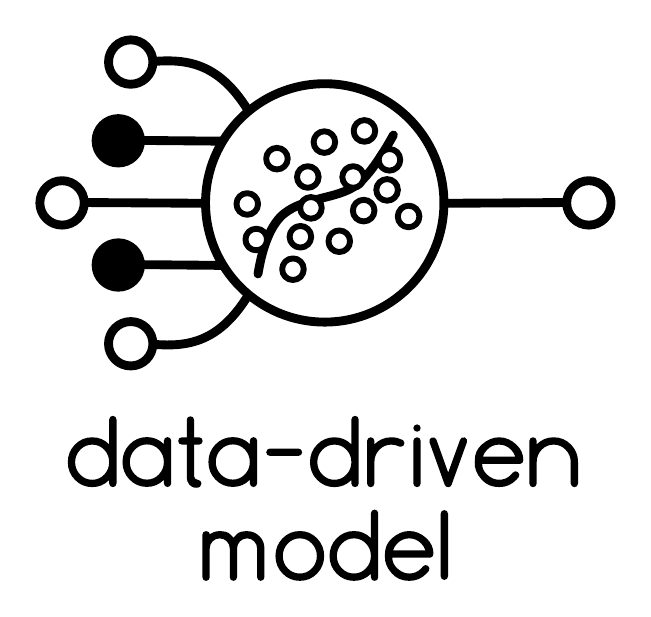}
\caption{Representation of a data-driven model. After receiving data from different sources, it infers a model for the data that can be used to  produce predictions.}\label{fig:icon_data_driven_modeling}
\end{figure}

Up till this point, we have described several techniques and scenarios in which substantial amounts of data are generated: the extraction techniques from Section \ref{sec:extractors} distill real-world and sensor observations into numerical information and mathematical representations; and, in section \ref{sec:blueprintITS}, Hypatia interacts with an ITS that relies on a database of student information and completed student tasks, which increases each time a student uses the system. In modern AI systems, data mining and machine learning techniques are used to analyze these data and to convert them into insights and practical models. These techniques, which we will refer to as \emph{data-driven modeling}, cover a broad area and make up the final class of AI in ME (Fig. \ref{fig:icon_data_driven_modeling}).

Data-driven modeling is employed for several reasons in ME. First, it may allow building models that are used to improve specific aspects of the learning process of individual students. These include AI models to predict a student's performance \cite{cortez2008using,smith2015diagrammatic,asif2017analyzing}, to determine at what specific problem step a student learns a concept \cite{baker2010detecting}, or even to detect that a student tries to game an ITS \cite{baker2008students} \footnote{We discuss student modeling in more detail in Section \ref{sec:narrow}.}. The ML techniques that are used to construct these models are mainly regression techniques (to obtain predictors that produce numerical values) and classification algorithms (to predict categorical or qualitative variables).

Second, data-driven modeling techniques can be used on large collections of student data, in a big-data fashion. A classic application in this category consists in analyzing completed student tasks in order to build a database of common errors, or ``bugs'' \cite{wenger1987artificial,Chrysafiadi2013}, which is an important component of an ITS. A different application consists in modeling complete student populations, which can be useful to group students into different ``stereotypes'', and is typically performed by clustering algorithms. Another application is the large-scale analysis of student profiles and completed tasks to improve the personalization of learning paths in an ITS. In this case, recommendation algorithms can be employed \cite{Chrysafiadi2013}. Finally, while studies in this field are mostly restricted to single schools or data from single ITS, it is easy to imagine that data-driven modeling can be applied to larger populations of students, for instance on a national level, where they could be used to make statistical assessments about the effectiveness of specific aspects of a curriculum.

\section{The present, revisited}
\label{sec:analysis}

Armed with the taxonomy introduced in Section \ref{sec:taxonomy}, we can now revisit the examples from Section \ref{sec:examples} and analyze their AI and ML techniques in more detail, pointing out some capabilities that may be added in the future.

\subsection{AI-based calculators}

\begin{figure}
\centering
\includegraphics[scale=0.2]{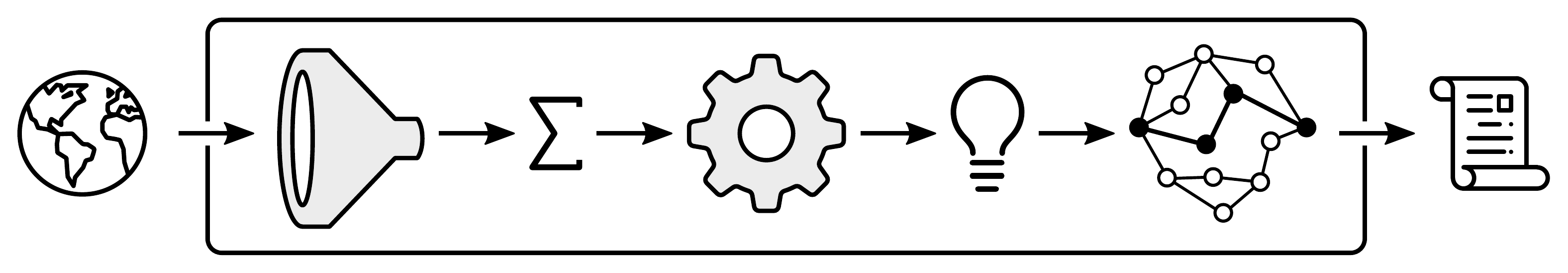}
\caption{The workflow of a camera calculator.}
\label{fig:reasonator}
\end{figure}

The ``camera calculator'', described earlier, operates as shown in Figure \ref{fig:reasonator}: First, the user captures a problem, for instance by taking a picture, which is translated by an information extractor into its mathematical formulation. Second, a reasoning engine solves the problem and produces a solution, in machine code. Third, an explainer translates the machine code into a human-readable reasoning sequence. In the case of simple problems, the reasoning engine and explainer could be replaced by a single module. Finally, the complete solution is presented to the user, who may request additional information on each of the steps.

The described workflow is valid for a wide range of calculators: The extractor could operate on different types of data, such as text from word problems, or voice commands, which are transcribed to text. As for the reasoning engine, many ATP and advanced computational engines are available, including the solvers Mathematica\footnote{\url{https://www.wolfram.com/mathematica/}} and  Maple\footnote{\url{https://www.maplesoft.com/}}. In this context, a pioneering role in the development of AI-based calculators is played by the WolframAlpha computational engine\footnote{\url{https://www.wolframalpha.com/}}, which operates on written queries and combines database look-ups with the computational power of Mathematica. It includes some \emph{explainer} capabilities as well, as it provide feedback on the solution and links to related educational resources. WolframAlpha was launched in 2009, making it a forerunner of current AI-based calculator such as the ones included in personal digital assistants, notably Siri\footnote{\url{https://www.apple.com/siri/}}, Cortana\footnote{\url{https://www.microsoft.com/cortana/}}, Alexa\footnote{\url{https://developer.amazon.com/alexa/}} and Google Assistant\footnote{\url{https://assistant.google.com/}}.

Another type of AI-based calculator is found in DGS with reasoning capabilities, which take geometric constructions as an input. While these tools contain a reasoning engine in the form of their GATP, they do not have explainer capabilities, as mentioned before, since the GATP they include produce proofs that cannot be translated to human-readable reasoning \cite{quaresma2020automated,kovacs2020geogebra}.

Presently, it is not clear how these new tools should fit in current mathematics curricula. If they are allowed without restrictions, some opponents claim that they will keep students from learning. Others recognize that the role of these tools must be debated in the educational community. Some proponents point out that the availability of these tools produces a shift in the desired objectives of ME \cite{kovacs2020geogebra}.

\subsection{Intelligent tutoring systems}
\label{sec:its}
An ITS is a computer-based learning tool that makes use of AI to create adaptive educational environments that respond both to the learner's level and needs, and to the instructional agenda \cite{graesser2012intelligent}.
While an ITS may share some underlying technologies with the AI-based calculators we described, they are much more complex tools, and they are fundamentally interactive. Here, we will review and discuss some relevant ITS that have been proposed in the literature.

\begin{figure}
\centering
\includegraphics[scale=0.2]{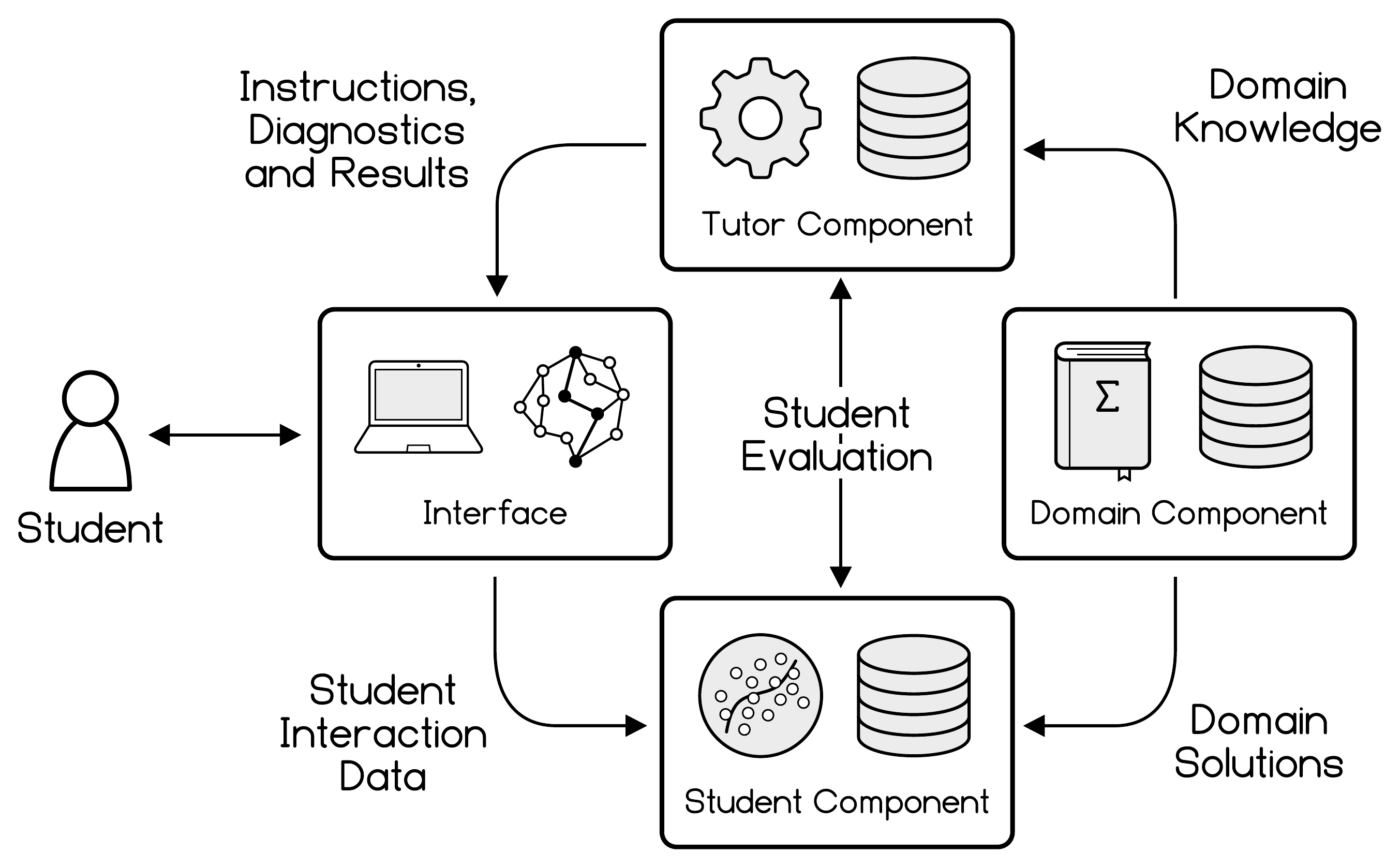}
\caption{Components of an ITS, in relationship to the introduced taxonomy.}
\label{fig:its}
\end{figure}

Typically, an ITS involves four different components \cite{wenger1987artificial,de2002designing,shute1994intelligent}, as represented in Figure \ref{fig:its}:
i) a domain component to encode the expert knowledge, ii) a student component to represent student knowledge and behavior, iii) a tutor component to select the best pedagogical action, and iv) an interface to interact with the student. The domain component includes, among others, expert knowledge, databases of tasks, and databases of bugs. The student component includes student models and student data, such as the detailed history of completed student tasks. After each interaction, the student actions are analyzed and the models are updated to reflect the new data. During the interaction with the student, the tutor uses a reasoning engine to track the reasoning of the student. It uses data from the student and domain components to spot bugs, offer feedback and personalize the learning path.

In some ITS, the tutor relies on expert knowledge with exact inference rules, which allow it to know a priori all possible solution paths to a problem. Examples include the Hypergraph Based Problem Solver (HBPS) ITS, which deals with word problems \cite{Arnau2013}, and the QED-Tutrix ITS, used for solving high-school level geometric proof solving problems \cite{leduc2016qed}. In general though, much of the information available in an ITS is incomplete or uncertain. Modeling the student, for instance, involves making inferences about the student's knowledge and behavior. Hence, probabilistic and approximate techniques such as Bayesian networks or fuzzy modeling are needed. An example is found in the TIDES ITS, proposed in \cite{danine2006tides}, which uses a Bayesian network to model student behavior based on the bugs that the student commits.

Currently, large parts of ITS that are being used in practice are designed and adapted using data-driven approaches, as described in Section \ref{sec:data_driven_modeling}. For instance, \cite{kurvinen2020long} describes the ViLLe ITS, whose commercial version uses AI techniques to improve the learning experience, based on the data of millions of student interactions.

\subsubsection{Narrow student modeling}
\label{sec:narrow}
The literature on student modeling is vast, and currently there exist dozens of student models that are used in practical ITS \cite{Chrysafiadi2013,sani2016artificial,abyaa2019learner}. Nevertheless, the majority of student models focus only on one specific aspects of the student, which is why we will refer to them as specific or ``narrow'' models. For instance, a student model may be constructed solely to predict student performance, and a different model may represent their competences in mathematics.
A comprehensive student model, as envisioned decades ago, should include both a \emph{complete} model of the student's knowledge as well as model of his behavior \cite{balacheff1993artificial}. This requires a more advanced AI, which we will discuss briefly in Section \ref{sec:studentmodeling}.

\subsubsection{Some notes on exploration, creativity and randomness}
\label{sec:exploration}

The interaction with an ITS guarantees that a specific learning occurs and that a target performance is reached. Nevertheless, if the ITS is to provide a rich experience in which it can ``determine the nature of the underlying meaning'', it should contain environments that allow the student to freely explore problem situations \cite{Balacheff1996}. This is obtained by ``guided discovery learning'', in which the system can shift between a tutor-like behavior for some situations and an open, exploratory environment for others.

In the AI literature, exploration is a prevailing theme. It is especially used in the subfield of ML known as ``reinforcement learning'', in which an agent explores an environment to determine how to maximize some reward over time. Successful applications of this field include robotics software, where the AI has to learn how to interact with the physical world, and strategy games, where the system must figure out how to beat the game using its custom set of rules.

AI is usually not associated with creativity, or the capability to come up with creative solutions. What is more, popular belief has it that AI is suited only to provide ``mechanical'' solutions, while creativity is reserved for humans. Nonetheless, it is precisely the AI techniques that require exploration that show strong indications that AI is capable of showing creative behavior. A striking example was seen in the 2015 tournament of the game Go between world champion Lee Sedol and AlphaGo, an AI-based Go program developed by Google DeepMind \cite{silver2017mastering}. The AI was first trained on records of human Go games, and then set out to battle clones of itself in order to continue improving. Interestingly, this approach led it to discover strategies that were previously unknown to human players.

In general, exploration and creativity require a certain component of ``randomness''. Several studies have been performed on this topic in the AI literature. For instance, randomness can be used to initiate the exploration of the solution space in neural networks. A comprehensive introduction to this topic can be found in \cite{scardapane2017randomness}.

\section{Modeling the mathematical learner: a most ambitious goal}
\label{sec:studentmodeling}

While the advances in AI and ML over the past decade have been impressive, it is important to put them in context. In particular, the recent well-known ``breakthroughs'' in AI are all techniques that are very good at solving a very specific problem, such as recognizing objects in pictures or translating text into a different language. If the specific setting is changed, they may not function properly. For instance, if a model is trained on recognizing animals in \emph{pictures}, it may not return a correct answer when given a \emph{drawing} of an animal. This capability of transferring a learned concept from one situation to a new context is known as ``generalization'', and humans are particularly good at it. It is also one of the traits expected from Artificial General Intelligence (AGI), as opposed to the described narrow AI systems. A discussion on generalization capabilities in AI can be found in \cite{kansky2017schema}.

In the previous sections, we have discussed several examples of student modeling, most of which are narrow modeling techniques, as they only cover one specific aspect of the student's knowledge or behavior. In order to design a complete student model, we believe the AI needed is akin to AGI, in that it should thoroughly understand several fields and it should be able to generalize. The following is a non-exhaustive list of properties: First, it should master the understanding of physics, which is being researched in robotics AI. In a mathematics learning setting, physics are often needed to interpret word problems, and they are indispensable to describe what is happening in photographic imagery. Second, it should feature strong natural language understanding. This is currently a very active field in ML, with the best results being obtained by large neural networks. One such noteworthy system is the Generative Pre-trained Transformer 3 (GPT-3), a neural network built by OpenAI that contains 175 billion parameters and was trained on 45 TB of text data \cite{brown2020language}. It shows remarkable text-analysis capabilities and can correctly answer many text queries by producing arbitrarily long human-like text passages. Third, the AI should have reasoning capabilities that allow it to solve mathematical and other problems. As briefly touched throughout this chapter, this would require cognitive abilities in mathematical, logical, and abstract reasoning. Its generalization capabilities would furthermore allow it to relate knowledge from different fields. Finally, it would require knowledge from cognitive and developmental psychology to understand the student's actions and general behavior. This aspect is perhaps the most complex to model, and ML research in this area is currently very limited.

\section{Conclusions and discussion}
\label{sec:conclusions}

In this chapter, we have presented an overview of contemporary AI techniques that are being used in digital ME tools. To provide a framework for this analysis, we have established a taxonomy of four different classes that cover each of these techniques: Information extractors, which convert data from the real world into a mathematical representation; Reasoning engines, which are solvers for mathematical problems; Explainers, which translate machine reasoning into human-interpretable steps; and data-driven modeling techniques, which are used to distill useful information and models from the data generated by students, for instance in ITS. 
We have also given more an in-depth analysis of AI-based calculator apps, which we consider to be the next generation of pocket calculators, and we have related the proposed taxonomy to the different components in a modern data-driven ITS.

We leave the reader with some ideas on AI-based tools for ME that we may see in the near future. First, progress in AI is currently dominated by ML-based techniques. The influence of these techniques is also noticeable in the experimental tools for ME that we have discussed. For instance, ML is used to automate perception in information extractors, to encode human intuition for searching in large solution spaces, and to analyze large volumes of data that are being generated in online education platforms. This trend will likely continue, with advances in ML being used to improve digital ME tools. %

Second, a recurrent theme throughout this chapter is the existence of parallels between research in AI and research in ME. For one, many of the questions asked in the design of digital ME tools can be found in AI research as well. And, as such, many state-of-the-art techniques that are developed in AI can solve problems that are encountered while building ME tools, in particular in ITS. Nevertheless, it must be noted that the mentioned AI techniques were developed in fields other than ME, with goals other than ME in mind, after which they were ``borrowed'' to be used in ME tools. This imbalance is certainly fueled by the massive interest that exists nowadays in the AI space, and one could wonder what incentives in research and industry would be required to start a new generation of ME-first AI techniques. A related observation is that the field of AI has advanced greatly over the past decade, partly due to the habit of publishing novel algorithms as open source software. In ME, digital tools are currently difficult to access: Most of them are either private (and closed-source) initiatives, or academic prototypes that are not maintained after their research project finishes. A notable exception is found in DGS such as GeoGebra.

Finally, we can reflect on the transformation that occurred around five decades ago with the advent of pocket calculators. At that time, there existed a large experimental space in which many different ideas were tried out, after which some standard tools emerged that are still in use today. Currently, AI-based ME tools are in a similar experimental phase, and although it may take some years or decades, we do expect to see a similar appearance of a set of standard AI-based applications for ME.

\bibliographystyle{unsrt}
\bibliography{bibliography}

\end{document}